\documentclass{nature}
\usepackage{lscape} 
\usepackage{lineno}
\usepackage{url}
\usepackage{graphicx}
\usepackage{amsmath}
\usepackage{adjustbox}
\usepackage{amssymb}
\usepackage{xcolor}
\usepackage{caption}

\newcommand{\citet}[2]{#1\cite{#2}}

\newcommand{\araa}{Annu. Rev. Astron. Astrophys.}   
\newcommand{\apj}{Astrophys. J.}   
\newcommand{\apjl}{Astrophys. J. Lett.}   
\newcommand{\apjs}{Astrophys. J. Suppl. Ser.}   
\newcommand{\aap}{Astron. Astrophys.}   
\newcommand{\mnras}{Mon. Not. R. Astron. Soc.}   
\newcommand{\nat}{Nature} 
\newcommand{\pasa}{Publ. Astron. Soc. Aust.}   
\newcommand{\sovast}{Soviet Astron.}   
\newcommand{\ssr}{Space Sci. Rev.}   

\makeatletter
\let\saved@includegraphics\includegraphics
\AtBeginDocument{\let\includegraphics\saved@includegraphics}
\renewenvironment*{figure}{\@float{figure}}{\end@float}
\makeatother

\title{Bow Shock and Local Bubble Plasma Unveiled by the Scintillating Millisecond Pulsar J0437$-$4715}

\author{Daniel J. Reardon$^{1,2}$, Robert Main$^{3,4,5}$, Stella Koch Ocker$^{6,7}$, Ryan M. Shannon$^{1,2}$, Matthew Bailes$^{1,2}$, Fernando Camilo$^{8}$, Marisa Geyer$^{9,8}$, Andrew Jameson$^{1}$, Michael Kramer$^{3}$, Aditya Parthasarathy$^{10,11,3}$, Ren\'ee Spiewak$^{1}$, Willem van Straten$^{12}$, Vivek Venkatraman Krishnan$^{3}$}

\begin{document}

\maketitle

\begin{affiliations}
 \item Centre for Astrophysics and Supercomputing, Swinburne University of Technology, P.O. Box 218, Hawthorn, Victoria 3122, Australia
 \item Australian Research Council Centre of Excellence for Gravitational Wave Discovery (OzGrav)
 \item Max-Planck-Institut f\"{u}r Radioastronomie, Auf dem H\"{u}gel 69, 53121, Bonn, Germany
 \item  Department of Physics, McGill University, 3600 rue University, Montr\'{e}al, QC H3A 2T8, Canada
 \item Trottier Space Institute, McGill University, 3550 rue University, Montr\'{e}al, QC H3A 2A7, Canada
 \item Cahill Center for Astronomy and Astrophysics, California Institute of Technology, Pasadena, CA 91101, USA
 \item The Observatories of the Carnegie Institution for Science, Pasadena, CA 91101, USA
 \item South African Radio Astronomy Observatory, Liesbeek House, River Park, Gloucester Road, Cape Town 7705, South Africa
 \item High Energy Physics, Cosmology \& Astrophysics Theory (HEPCAT) Group, Department of Mathematics \& Applied Mathematics, University of Cape Town, Cape Town 7700, South Africa
 \item ASTRON, Netherlands Institute for Radio Astronomy, Oude Hoogeveensedijk 4, 7991 PD Dwingeloo, The Netherlands
 \item Anton Pannekoek Institute for Astronomy, University of Amsterdam, Science Park 904, 1098 XH Amsterdam, The Netherlands
 \item Manly Astrophysics, 15/41-42 East Esplanade, Manly, NSW 2095, Australia
\end{affiliations}


\begin{abstract}
%
%
%
%

The ionized interstellar medium contains au-scale (and below) structures that scatter radio waves from pulsars, resulting in scintillation. Power spectral analysis of scintillation often shows parabolic arcs, with curvatures that encode the locations and kinematics of the pulsar, Earth, and interstellar plasma.
Here we report the discovery of $\mathbf{25}$ distinct plasma structures in the direction of the brilliant millisecond pulsar, PSR~J0437$\mathbf{-}$4715, in observations obtained with the MeerKAT radio telescope.
Four arcs reveal structures within $\mathbf{5000}\,$au of the pulsar, from a series of shocks induced as the pulsar and its wind interact with the ambient interstellar medium. The measured radial distance and velocity of the main shock allows us to solve the shock geometry and space velocity of the pulsar in three dimensions, while the velocity of another structure unexpectedly indicates a back flow from the direction of the shock or pulsar-wind tail.
The remaining $\mathbf{21}$ arcs represent a surprising abundance of structures sustained by turbulence within the Local Bubble -- a region of the interstellar medium thought to be depleted of gas by a series of supernova explosions about $\mathbf{14}\,$Myr ago. The Local Bubble is cool enough in areas for sub-au density fluctuations to arise from turbulence.
\end{abstract}

Energetic processes in our Galaxy shape its evolution by fueling star formation and driving turbulence in the interstellar medium that induces a cascade of density irregularities in the interstellar plasma\cite{Armstrong+95, Goldreich+95}. Radio waves from pulsars are scattered off the resulting small-scale ($\lesssim$\, au) plasma structures and subsequently interfere to produce frequency-dependent interference patterns that drift across our radio telescopes with a velocity that depends on the relative motions of the pulsar, Earth, and plasma. The measured intensity variations (known as scintillation\cite{Rickett69, Cordes+86, Rickett90}) can reveal parabolic ``scintillation arcs" in their power spectra, with curvatures that encode the velocities and locations along the line-of sight of individual plasma structures.

A pilot survey for scintillation arcs towards PSR~J0437$-$4715 was conducted with the 64-antenna MeerKAT radio telescope shortly after it commenced scientific operations\cite{Bailes+20}. This observation spanned 10.5\,hours, in a frequency band from 856\,MHz to 1712\,MHz with 4096 frequency channels. A dynamic spectrum of the observation shows the scintillation as a function of frequency and time (Figure 1). A power spectrum of the observed intensity variations (the secondary spectrum) reveals that the power is organised into multiple parabolic arcs (Figure 2). The observation showed evidence for at least $25$ of these scintillation arcs, caused by scattering from turbulent plasma along the line of sight. The arc curvature depends on the distance, velocity, and orientation of the scattering layer relative to the pulsar\cite{Stinebring+01, Walker+04, Cordes+06}, with multiple arcs at independent curvatures implying multiple scattering layers. Scintillation studies can even probe the local environments of pulsars, such as supernova remnants\cite{Yao+21, Serafin+24} and companion-star atmospheres\cite{Lin+23}. The arcs we report here include extremely low-curvature arcs (inset of Figure 2), which as we describe below, originate from the pulsar bow shock\cite{Bell+1993}. The number of arcs is unprecedented and provides a deep view into the warm ($\sim10^4$ K) plasma interior of the Local Bubble.

\begin{figure}
\centering
  \includegraphics[width=.8\textwidth]{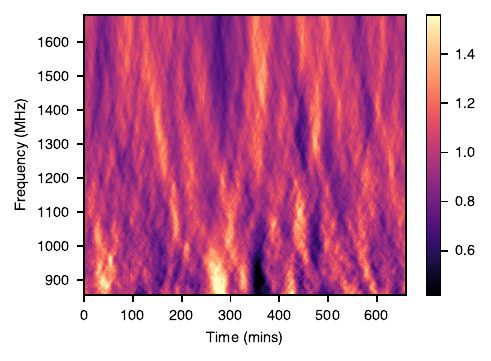}
  \caption{Dynamic spectrum of intensity from a MeerKAT observation of PSR J0437$-$4715 on MJD 58843. The apparent intensity variations with time and observed radio frequency are due to scintillation. The colour scale shows intensity on a linear scale, normalised by the mean. The pulsar is observed in weak scintillation, where the intensity is not fully modulated, and becomes weaker at higher frequencies.}
\label{fig:svd}
\end{figure}

\begin{figure}
\centering
  \includegraphics[width=0.9\textwidth]{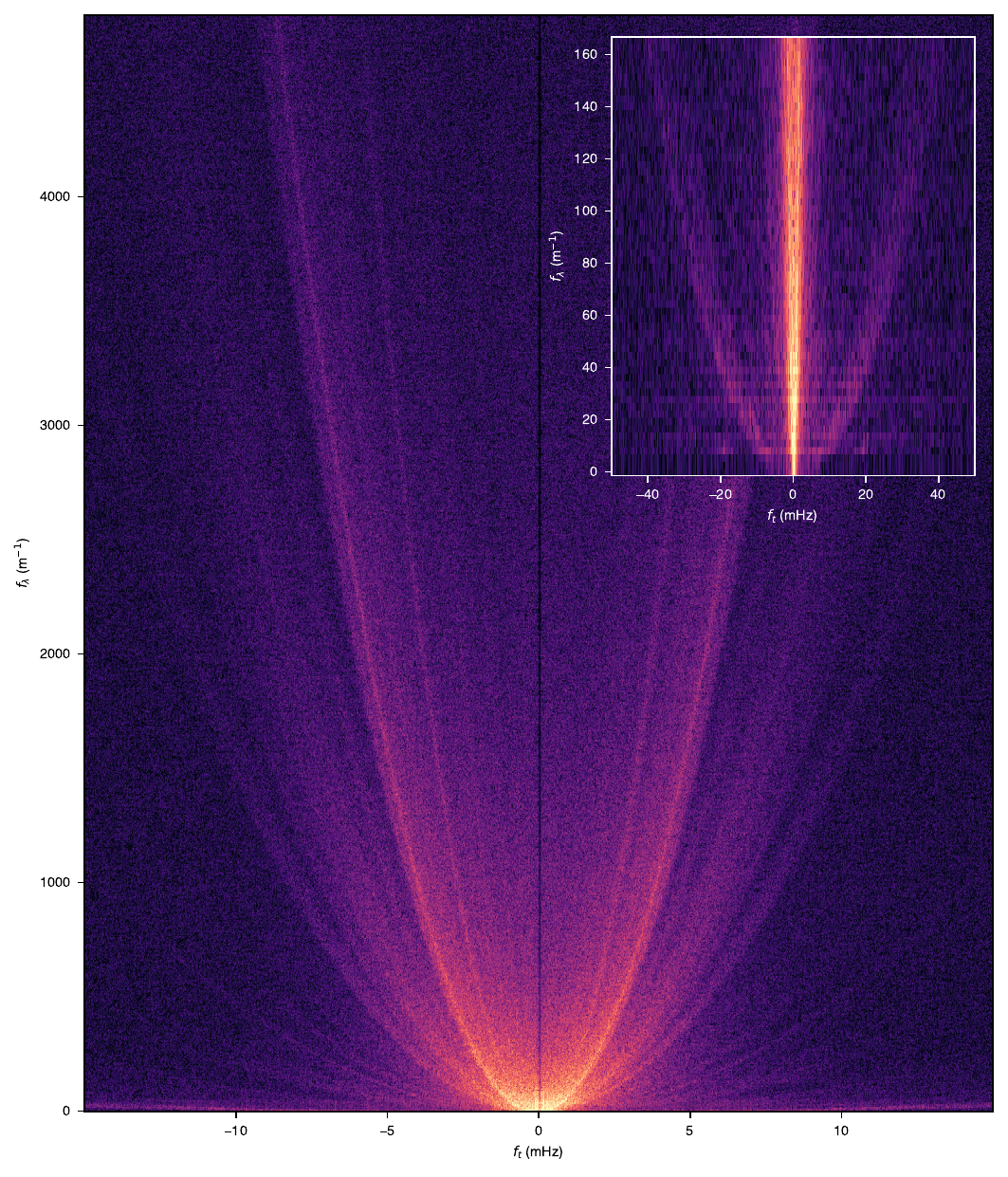}
  \caption{Secondary spectrum from a MeerKAT observation of PSR J0437$-$4715 on MJD 58843. The plot shows the power spectrum of the intensity fluctuations as a function of Fourier conjugates of time $f_t$ and wavelength $f_\lambda$. This observation shows evidence for more than 20 high curvature scintillation arcs from the interstellar medium and four low-curvature bow shock scintillation arcs (inset). The colour shows power on a logarithmic scale (with a range of 62 dB). An animated version of this figure is available in the online supplementary material and shows the variations in the arcs arising from the changing orbital velocity of the pulsar.}
\label{fig:sspec}
\end{figure}

The curvatures of the scintillation arcs vary with the inverse square of the transverse velocity of the line of sight through the plasma, $\mathbf{V}_{\rm eff}$\cite{Reardon+20}. As this pulsar is in a $5.7$-day orbit with a white-dwarf companion\cite{Johnston+93}, a full scintillation arc survey was conducted over six days to measure the changes in the orbital velocity of the pulsar that are imprinted in the arc curvatures. Scattering screens identified over multiple days were modelled to measure their precise distances and velocities (see methods), which are summarised in Table 1. The arc curvature variations and inferred models are shown in Figure 3. Such precise models are possible from a single orbit in part because precision pulsar timing of PSR~J0437$-$4715 has already determined the orbital geometry in three dimensions\cite{vanStraten+01} and the pulsar distance to high precision, $D=156.96 \pm 0.11$\,pc\cite{Reardon+24}. We present the distance and velocity models for $21$ scintillation arcs originating from interstellar plasma density fluctuations and four that we attribute to structures within the pulsar bow shock.

\begin{table}
\label{tab:params}
\centering
\caption{Estimated scattering screen properties. $\gamma$ is the spectral index of the power distribution of the arc and $\tau_{\rm max}$ is the maximum extent in differential time delay, $f_t$. The velocities for the 21 interstellar scintillation arcs are with respect to the Solar system barycentre, while the four bow shock arcs are with respect to the pulsar. For each measurement, the numbers in parentheses are the uncertainties on the last quoted decimal place. The screens are sorted by their distance from Earth, as determined by the preferred geometric model for each screen. The log Bayes factor in support of the anisotropic model over isotropic is $\ln \mathcal{B}$. The velocities are in units of km\,s$^{-1}$. }
\medskip
{\linespread{1.4}\selectfont
\begin{adjustbox}{width=1\textwidth}
\begin{tabular}{|c|c|c|ccc|ccc|c|}
\hline
 & & & \multicolumn{3}{c|}{Isotropic} & \multicolumn{3}{c|}{Anisotropic} &   \\
\hline
ID & $\gamma$ & $\tau_{\rm max}$ ($\mu$\,s) & $s$ & $V_{{\rm IISM},\alpha}$  & $V_{{\rm IISM},\delta}$& $s$ & $\zeta$ ($^\circ$ N$\rightarrow$E)  & $V_{{\rm IISM},\zeta}$ & $\ln \mathcal{B}$  \\
\hline
\hline
\footnotesize
1 & 3.42(14) & 1.192  & 0.468(17) & -33(3) & 22(5) & 0.471(17) & 127(3) & -41.0(15) & 1.5 \\ 
2 & 3.63(15) & 1.030  & 0.472(7) & 59(3) & 40.7(6) & 0.469(4) & 178(2) & -38(2) & 8.3 \\ 
3 & 3.43(18) & 1.597  & 0.442(6) & -24(2) & 21(3) & 0.445(3) & 131(2) & -33.8(5) & 13.0 \\ 
4 & 3.58(8) & 1.048  & 0.33(3) & 42(7) & 44(5) & 0.33(3) & 164(5) & -30(7) & 2.2 \\ 
5 & 3.38(14) & 0.871  & 0.323(5) & -14(3) & 25(3) & 0.325(6) & 131(2) & -27.6(12) & -0.6 \\ 
6 & 3.79(15) & 0.605  & 0.289(16) & 83(2) & -7.8(15) & 0.289(19) & 19(4) & 21(6) & 0.8 \\ 
7 & 3.81(16) & 0.639  & 0.28(3) & 90(5) & 2(3) & 0.27(3) & 13(8) & 23(9) & 2.5 \\ 
8 & 3.72(14) & 0.610 & 0.1208(19) & 88.8(4) & -39.5(3) & 0.251(5) & 27(3) & 18(4) & 6.1 \\ 
9 & 3.27(18) & 0.457  & 0.183(10) & 30(5) & 46(4) & 0.178(8) & 148(3) & -23(4) & 4.3 \\ 
10 & 3.03(18) & 0.549  & 0.152(5) & 14(4) & 47(3) & 0.154(5) & 146(2) & -33(2) & -0.8 \\ 
11 & 3.1(2) & 0.546  & 0.146(5) & 63(4) & 49(2) & 0.145(4) & 161(3) & -27(3) & 3.4 \\ 
12 & 3.0(2) & 0.461  & 0.087(3) & 39(4) & 40(3) & 0.086(2) & 155(2) & -21(3) & 6.6 \\ 
13 & 3.1(2) & 0.319  & 0.082(6) & 11(4) & 27(5) & 0.081(3) & 139(2) & -13(3) & 7.5 \\ 
14 & 3.16(16) & 0.480  & 0.079(7) & 51(4) & 29(4) & 0.076(5) & 158(3) & -7(4) & 5.0 \\ 
15 & 3.0(2) & 0.221  & 0.0686(11) & 0(3) & 21(4) & 0.069(3) & 124(2) & -13(2) & -9.8 \\ 
16 & 2.6(3) & 0.052  & 0.0230(12) & 51(5) & 48(4) & 0.0226(11) & 155(3) & -21(4) & 3.3 \\ 
17 & 2.7(3) & 0.077  & 0.0226(14) & 81(5) & 41(3) & 0.0218(12) & 173(3) & -30(5) & 2.2 \\ 
18 & 2.3(2) & 0.044  & 0.0148(14) & 36(5) & 31(5) & 0.0152(12) & 153(3) & -14(6) & 3.4 \\ 
19 & 1.5(3) & 0.026  & 0.00441(18) & 67(4) & 28(2) & 0.0045(3) & 162(3) & -8(5) & -3.1 \\ 
20 & 1.3(3) & 0.015  & 0.00339(17) & 65(5) & 46(3) & 0.00328(10) & 167(2) & -29(4) & 7.7 \\ 
21 & 1.0(4) & 0.010  & 0.00(2) & 20(3) & 30(3) & 0.0012(3) & 143(5) & -10(12) & 1.5 \\ 
\hline
\hline
Shock A & 1.28(14) & 0.047  & 0.000148(9) & -47(3) & 54(3) & 0.000146(6) & 141(3) & $-$72(3) & 8.8 \\ 
Shock B & 0.5(3) & 0.005  & 0.000114(15) & 14(3) & -19(4) & 0.000108(14) & 133(4) & $-$117(7) & 2.6 \\ 
Shock C & 1.7(5) & 0.018  & 0.0(3) & -88(6) & 82(7) & 0.000107(5) & 139(2) & 26(2) & 11.0 \\ 
Shock D & 1.31(17) & 0.033 & 0.000086(4) & 33.3(18) & 35.8(19) & 0.000084(4) & 139(3) & $-$50(3) & 2.2 \\ 




\hline
\end{tabular}
\end{adjustbox}}
\end{table}

\begin{figure}
\centering
  \includegraphics[width=1\textwidth]{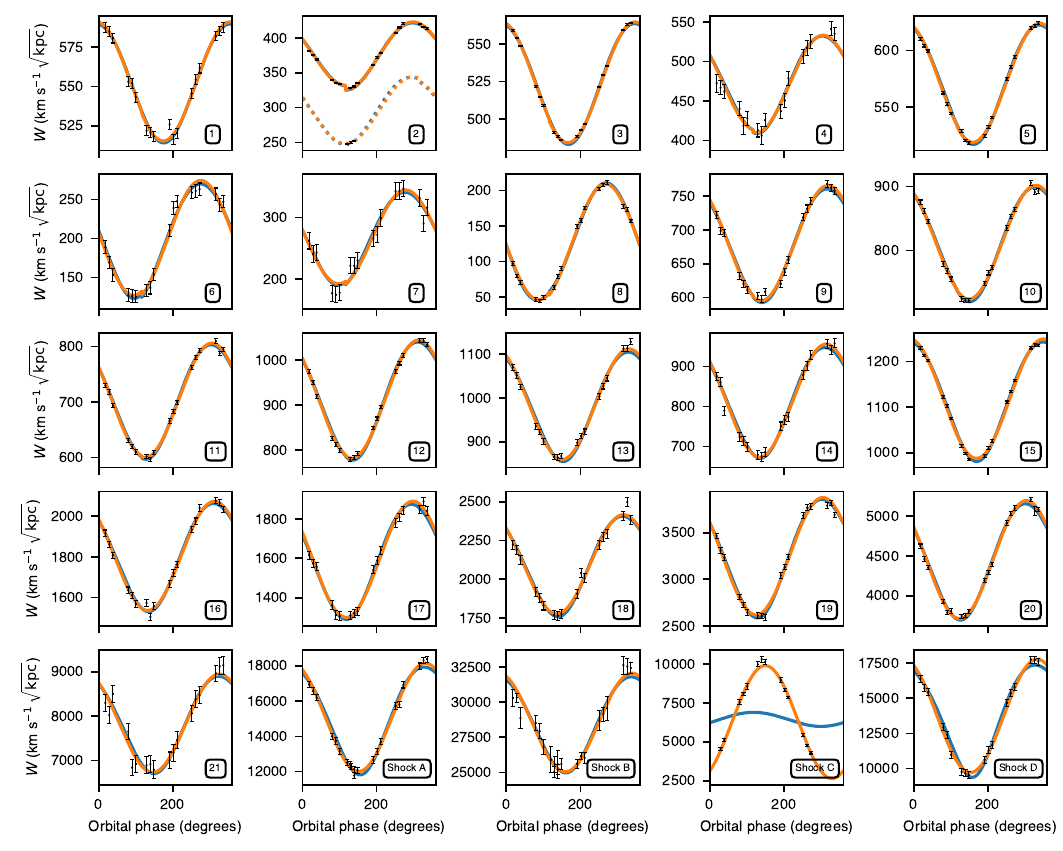}
  \caption{Measurements and models for all of the scintillation arcs described in this work. Each sub-panel, labelled in the lower right with the arc ID from Table 1, shows the $W$ measurements (mean) and uncertainties (standard deviation), $\sigma_{\rm new}$ (modified using the white noise parameters inferred from the best model; see methods), from one scintillation arc as a function of the orbital phase of the pulsar. The measured $W$ values are related to the curvature of the scintillation arc, $\eta$ with $W = 1/\sqrt{2\eta}$. The inferred models for isotropic and anisotropic scattering screens are shown in blue and orange lines, respectively. For arc ID 2, the measurements from the earlier October observation are shown, with the corresponding models for that epoch given in dashed lines. The small model discontinuities that are visible for this arc, and others, are due to the observations phase wrapping during the six day observing campaign. The pulsar was observed at $\sim 130^\circ$ orbital phase for the first observation, causing the observations on the last two days to phase wrap and ending the full orbit with a discontinuity caused by the change in the velocity of the Earth, $\mathbf{V}_{\rm E}$, over the $5.74$ day orbital period of the pulsar.}
\label{fig:all_arcs} 
\end{figure}

Bow shocks are formed if the space velocity of an object (with an energetic wind) exceeds the magnetosonic (sound) speed of the interstellar medium (ISM). Many pulsars travel at such velocities, including PSR~J0437$-$4715 with a transverse velocity $\mathbf{V}_{\rm psr} = 104.7\,$km$\,$s$^{-1}$ relative to the solar system barycentre. Yet the sample of known pulsar bow shocks remains small because they are Balmer-dominated and require the presence of neutral gas to be observed\cite{Brownsberger+14}. As material from the ambient ISM is shocked, charge transfer and collisional excitation of the neutral Hydrogen yield observable emission from the H$\alpha$ Balmer transition. In some cases, including for PSR~J0437$-$4715, far ultraviolet (FUV) emission from the shocked ISM is also observed\cite{Rangelov+16}. This emission traces material behind the ``forward shock": the outer-most boundary that separates the ambient ISM from the shocked ISM. The boundary between the shocked ISM and the shocked pulsar wind is referred to as the ``contact discontinuity", while the inner-most boundary separating the pulsar wind from shocked wind is the ``termination shock" (or ``reverse shock"). X-ray and radio emission can arise between the termination shock and contact discontinuity. These boundaries separate the notable structures in a simple bow shock model. A schematic of the shock and these structures is shown in Figure 4, using an analytical expression for the shape of the contact discontinuity from \citet{Wilkin (1996)}{Wilkin96}. This model for the contact discontinuity is consistent with the observed H$\alpha$ emission near the head of the bow shock, whereas the H$\alpha$ emission departs from the model in the tail of the shock because of the post-shock pressure\cite{Bucciantini2002}.

\begin{figure}
\centering
  \includegraphics[width=.8\textwidth]{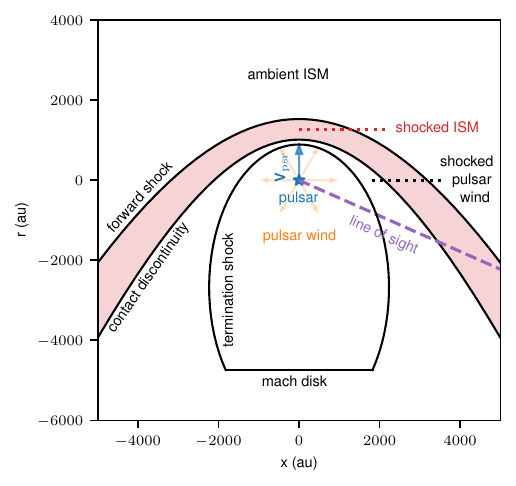}
  \caption{ Schematic of the PSR~J0437$-$4715 bow shock system, depicting the main boundaries that we propose are associated with the scintillation arcs in the MeerKAT observations. H$\alpha$ emission arises from the shocked ISM, which is bounded by the forward shock and the contact discontinuity shown here to scale and described by our fitted hyperbola model. The termination shock closes at a mach disk at an unknown distance behind the pulsar.}
\label{fig:shock}
\end{figure}

The bow shock of PSR~J0437$-$4715 was first observed in H$\alpha$ in 1993 and has an estimated transverse distance to the forward shock of $\sim 1400$\,au from the pulsar\cite{Bell+1993}. The bow shock was observed again in 2012 in H$\alpha$ (Figure 5) and FUV\cite{Brownsberger+14, Rangelov+16}. Unlike most other pulsars\cite{Brownsberger+14}, the bow shock is smooth and has remained stable in morphology over 20 years. The shock morphology suggests a near-isotropic pulsar wind and that the ISM local to the pulsar ($<10^4$ au) has approximately uniform density and constant velocity\cite{Vigelius+2007}. We therefore assume a simple geometric shock model constrained by both the combined H$\alpha$ emission and our scintillation arcs.

The curvature of a scintillation arc also depends on the distance to the scattering material, with scattering screens nearer to the pulsar or observer resulting in lower curvatures. From our analysis, we find distances that place the four lowest curvature (i.e., most open) arcs in the region of the pulsar bow shock. The brightest bow shock arc (Shock A in Table 1), originates from scattering that occurs $D_A=4720\pm190\,$au from the pulsar, radially along the line of sight. We attribute this arc to the shocked ISM traced by H$\alpha$ and FUV emission. Our model of the curvature variations across the orbit shows that an anisotropic distribution of plasma in Shock A is moving at $\Delta V_A = -72 \pm 3\,$km$\,$s$^{-1}$ relative to the pulsar, along an axis that is aligned at position angle $141^\circ \pm 3^\circ$ (East of North). This anisotropy is misaligned from the proper motion direction ($120.5^\circ$, East of North) by $\sim 20^\circ$. The Shock A arc is also broad, with power spanning a range of curvatures. Attributing this breadth to the radial thickness of the scattering layer implies a thickness $\mathcal{O}(1000)\,$au, which may correspond to the depth of shocked ISM along the pulsar LOS. A second, weaker, arc is also observed within this broad distribution of power. This arc originates from nearer to the pulsar at $D_D=2700\pm130\,$au, suggesting that it may be associated with the contact discontinuity, or second structure within the shocked ISM material. The plasma causing this arc (Shock D in Table 1, named according to the distance from the pulsar) has a lower velocity than Shock A, of $\Delta V_D = -50 \pm 3\,$km$\,$s$^{-1}$ relative to the pulsar.

 Using the projected transverse distance to the H$\alpha$ emission and our measured radial distance to the shocked material associated with Shock A, we have reconstructed a three-dimensional view of the bow shock. Our model is a hyperboloid (see methods and Figure 5), which is appropriate for a near-isotropic pulsar wind and a uniform-density ISM. The model follows the H$\alpha$ emission closely near the shock front and remains accurate at the position angles required for accuracy at the position of the pulsar, before departing in the tail\cite{Bucciantini2002}. We find that the shock is tilted away from the line of sight with an inclination of $i_b = 113 \pm 3^\circ$ (where $i=0^\circ$ describes a shock pointing towards the observer). H$\alpha$ observations with sufficient spectral resolution should be able to independently infer this inclination angle\cite{Romani+2022, Ocker+24}. Additionally, the position angle from the pulsar to the vertex of the shock front is misaligned by only $4.3\pm0.6$\,degrees from the proper motion. This position angle offset could be explained by a modest velocity of the ambient ISM of $\sim 8\,$km\,s$^{-1}$ but could also be due to unresolved anisotropy in the pulsar wind\cite{Wilkin00}. Assuming that the inclination of the shock front is dominated by the space velocity of the pulsar, this corresponds to a pulsar radial velocity of $V_r = -45\pm4\,$km\,s$^{-1}$. This radial velocity away from the observer (with respect to the solar system barycentre) would lead to an apparent second spin-frequency derivative and time derivative of the proper motion due to a geometric effect\cite{Shklovskii70, vanStraten03, Liu+18}, which will soon impact the precision timing observations of this pulsar\cite{Reardon+24}, and studies of nanohertz-frequency gravitational waves\cite{NGgw, PPTAgw, EPTAgw, CPTAgw} if it is not accurately modelled. 

\begin{figure}
\centering
  \includegraphics[width=.8\textwidth]{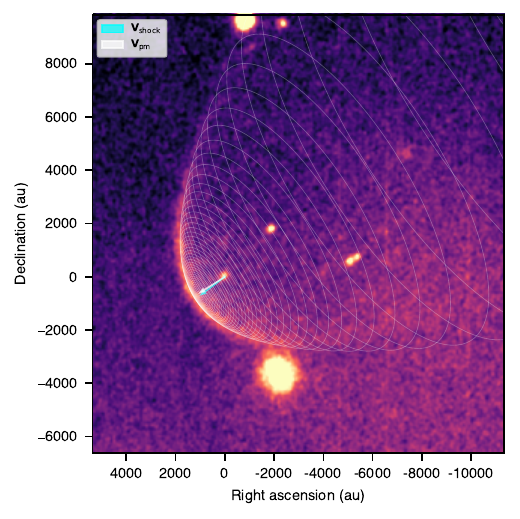}
  \caption{H$\alpha$ image of the bow shock of PSR~J0437$-$4715 with our hyperboloid model of the shocked ISM behind the forward shock. The data are from the Southern Astrophysical Research (SOAR) telescope\cite{Brownsberger+14}. The hyperboloid model is tilted away from a side-on view by $23^\circ$, such that we view slightly from the back. The white arrow indicates the direction of the proper motion and the blue arrow shows the best-fit direction for the vertex of the shock front. The white-dwarf companion of PSR~J0437$-$4715 is seen as the bright point at the origin of the arrows. Although the pulsar is not visible in the image, its position is indistinguishable from the companion star on this scale.}
\label{fig:shock}
\end{figure}

We observe two further, much weaker, bow shock arcs that are both closer to the pulsar than Shock A discussed above, and are thus likely associated with internal structures of the bow shock. The third bow shock arc (Shock B in Table 1) is the faintest and lowest curvature and is measured to be at a radial distance $D_B=3490 \pm 450\,$au, with a fast velocity $\Delta V_B = -117\pm7\,$km$\,$s$^{-1}$ relative to the pulsar. A fourth bow shock arc (Shock C in Table 1), observed mostly at higher curvatures to Shock A, varies in anti-phase with most of the other arcs, increasing in curvature while all other arcs decrease in curvature and vice versa. It is partially obscured by higher-curvature interstellar arcs during some observations. By modelling the curvature variations on the epochs where the arc is distinct, we have found that the arc originates from an anisotropic plasma structure, at $D_C=3460 \pm 160\,$au, and has a much lower velocity with respect to the pulsar of $\Delta V_C = 26\pm2$\,km\,s$^{-1}$ along the anisotropy, relative to the pulsar. The velocity is positive in the direction of the pulsar proper motion unlike the bulk flow responsible for arcs A, B, and D, which in the frame of the pulsar is rapidly streaming in the opposite direction. It therefore describes either a plasma back-flow, or plasma with a high velocity in the x-direction of Figure 4 such that the component transverse to the line of sight points towards the shock front. These shock measurements represent the first tomographic view of the dynamic internal layers of a pulsar bow shock.

Pulsar scintillation is a useful probe of structures in the ionized ISM that would otherwise be invisible on the ($\lesssim 1\,$mas) angular scales. The scintillation arcs can arise as a result of diffractive or refractive scattering in the ISM\cite{Cordes+86}, caused either by a turbulent cascade of density fluctuations\cite{Armstrong+95} or other discrete structures like corrugated reconnection current sheets\cite{Pen+14}. The remaining 21 modelled scintillation arcs originate from small-scale ($\lesssim$au) inhomogeneities in the interstellar plasma. The curvature of these arcs reveal modulation due to the change in the velocity of the Earth between our pilot survey and our orbital campaign two months later. This can be used to precisely constrain the distance to the pulsar. We measure $D_{\rm psr}=158.1\pm1.5$\,pc, which is of comparable precision to, and consistent with, the pulsar distance derived from very-long-baseline interferometry ($D_{\rm psr}=156.3\pm1.3$\,pc\cite{Deller+08}), but was achieved using a much shorter time baseline. However, our distance constraint utilises a measurement of the sky orientation of the pulsar orbit that was obtained from pulsar timing. If we assume no such prior knowledge, which is typical for other pulsars, we can measure the orbital orientation using the most sensitive scintillation arc, $\Omega=204^{+10}_{-13}$ (degrees, East of North to the ascending node), and obtain $D_{\rm psr}=158^{+3}_{-6}$\,pc (Extended Data Figure 1). To achieve precise results for other systems, scintillation arc curvatures can be monitored over a year to leverage a larger fraction of the Earth's orbit.

For PSR~J0437$-$4715, the arc morphology follows closely that expected of Kolmogorov turbulence\cite{Reardon+20}. However, we also observe small asymmetries in the arc curvatures as the apex of the arcs shift slightly from the origin, due to small ($\mathcal{O}(0.1)\,$mas) angular displacements of the pulsar image\cite{Cordes+06}. Since these asymmetries are shared between arcs and occur on short timescales, we believe they most likely originate from the ionosphere\cite{Yeh+82} (see methods). We have also quantified the strength of scattering by the plasma associated with each arc. Although the interstellar scintillation arcs appear to have a wide range of strengths in the secondary spectrum because of their varied distances, they are caused by similar plasma scattering angles. However, the scintillation arcs attributed to the bow shock originate from significantly stronger scattering (Extended Data Figure 2; see methods). Our observations demonstrate that discrete structures in the turbulent interstellar medium are ubiquitous. Indeed more interstellar scintillation arcs are likely present but are too faint to measure reliably at multiple epochs. At curvatures higher than that of the brightest arc in our secondary spectra, individual arcs are difficult to identify because of increased background power. This difficulty may explain why all of the arcs that we have modelled (Table 1) are located closer to the pulsar than to the Earth, with the nearest to Earth at a distance of $82.9\,$pc. 

Our observed number of arcs places a limit on the interstellar thin screen number density of $\rho_N\geq 0.14\,$pc$^{-1}$ on this line of sight, which is an order of magnitude higher than previous estimates from scintillation arcs\cite{Mckee+22, Ocker+24}. However, this is a conservative lower bound because of selection effects that determine which scintillation arcs are measurable in our spectra; notably a bias for arcs with lower curvature, or higher $\mathbf{V}_{\rm eff}$. If we assume that an unbiased sample of screens has no preferred velocity direction or screen location, then the inferred screen number density is $\rho_N\sim 1\,$pc$^{-1}$. While stability of the bow shock and scintillation arc morphology with time\cite{Brownsberger+14, Reardon+20} suggest a homogeneous ISM on the scale of hundreds of au (traversed by the pulsar over decades), significant inhomogeneities appear at parsec scales. 

According to the latest three-dimensional model of the Local Bubble\cite{Pelgrims+20, Zucker+22}, the edge of this region as traced by neutral gas and dust\cite{Lallement+19} is at a distance of 287\,pc in the direction of PSR~J0437$-$4715. The pulsar lies comfortably within this region, making it an unambiguous tracer of plasma structure internal to the Bubble. Dust extinction maps and velocity reconstructions of stars in local molecular clouds support a picture in which the Local Bubble was produced by a series of $\sim 15$ supernovae over the past $\sim 14$ Myr, which drove an expanding shell of dust and gas that triggered star formation along the edge of the Bubble\cite{Zucker+22}. However, the internal plasma properties of the Bubble remain highly uncertain, with lines of sight to nearby pulsars indicating mean electron densities that are over $10\times$ greater than expected for the million-degree gas thought to pervade the Bubble\cite{Spangler+2009, Linsky+2021}. 

The external morphology of the Bubble traced by dust contains significant inhomogeneities, possibly due to turbulence and the nascent structure of the ISM prior to the formation of the Bubble\cite{Kim+2017}. The prevalence of arcs observed for PSR~J0437$-$4715 indicates an extreme number density of warm plasma screens, demonstrating that the interior of the Bubble has cooled in areas to sustain highly turbulent density fluctuations down to sub-au scales (Extended Data Figure 3). There is additional support for cooler gas (neutral and partly ionized) in the Local Bubble from absorption spectra\cite{Welsh+10}. Some of these plasma screens may have formed with the Bubble itself, from interactions between successive supernovae and the ISM.

The discovery of an unprecedented number of plasma structures within the Local Bubble and a pulsar bow shock marks the exploration of the ionized ISM along the entirety of the line-of-sight towards PSR J0437$-$4715. The pulsar, as one stellar remnant, has enabled unique tomographic views of the plasma stirred by the energetic echoes of stellar processes in the solar neighbourhood. These results constrained the pulsar distance to $\sim$parsec precision, implying that such tomography could be conducted along the sight lines to pulsars without known distances.

\clearpage

\renewcommand{\figurename}{Extended Data Figure} 
\setcounter{figure}{0}

\begin{figure}
\centering
  \includegraphics[width=.8\textwidth]{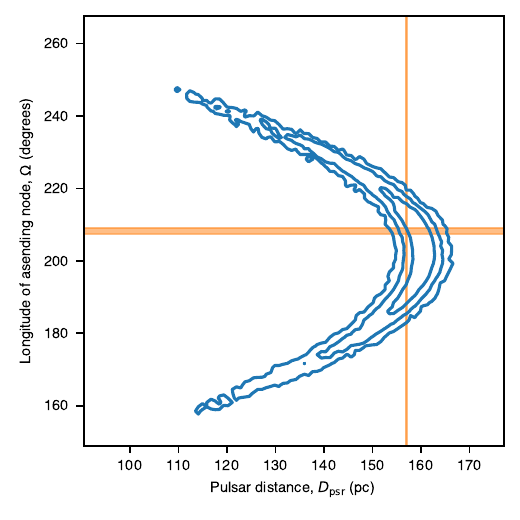}
  \caption{Probability density contours for the pulsar distance, $D_{\rm psr}$ and longitude of ascending node, $\Omega$, derived from a high signal-to-noise scintillation arc. Blue contours show the 68\%, 95\%, and 99.7\% credible intervals for the posterior distribution inferred from the arc (``ID 2" in Table 1). The orange shaded regions show the precise $1\sigma$ (68\%) confident intervals from pulsar timing\cite{Reardon+24}. The central 68\% credible intervals inferred from the scintillation arc are $\Omega=204^{+10}_{-13}$ (degrees, East of North) and $D_{\rm psr}=158^{+3}_{-6}$\,pc.}
\label{fig:kom-D}
\end{figure}

\begin{figure}
\centering
  \includegraphics[width=.8\textwidth]{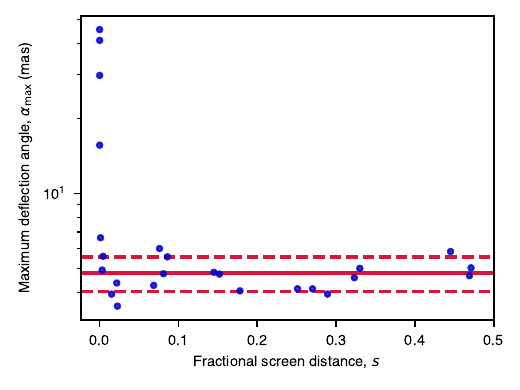}
  \caption{Strength of scattering for each screen. The maximum observed scattering angle ($\alpha_{\rm max}$) is shown in blue for each screen with fractional screen distance ($s$). The solid and dashed horizontal lines show the mean and standard deviation of the 
  $\alpha_{\rm max}$ values from screens attributed to interstellar plasma, $\alpha_{\rm max, ISM} = 4.8 \pm 0.8\,$ma
  s. The four bow shock scintillation arcs originate from significantly larger scattering angles.}
\label{fig:refraction}
\end{figure}

\begin{figure}
\centering
  \includegraphics[width=.9\textwidth]{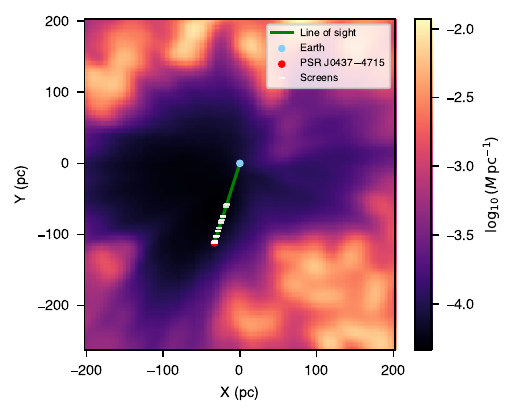}
  \caption{Location of the measured scattering screens along the line of sight to PSR J0437$-$4715, projected onto the Galactic plane. The colour scale shows the \citet{Lallement et al. (2019)}{Lallement+19} map of logarithmic differential extinction due to dust, $\log_{10}(A_v)$, where $A_v$ is in units of magnitude per parsec ($M\,{\rm pc}^{-1}$). The darker regions are used to define the latest (three-dimensional) Local Bubble model\cite{Pelgrims+20}.}
\label{fig:distance}
\end{figure}

\clearpage

\begin{methods}

The increased sensitivity of this scintillation arc study compared with previous observations\cite{Reardon+20} was enabled by the MeerKAT radio telescope, which consists of 64 offset Gregorian antennas, each with a diameter of  13.5\,m (nominal; 13.9\,m effective diameter). The resolution achieved in the secondary spectrum is determined by the observing time span and observing bandwidth. The MeerKAT L-band receiver has a wide bandwidth, from 856\,MHz to 1712\,MHz, which provides the excellent resolution that was key in identifying scintillation arcs from the pulsar bow shock. The observations were processed and recorded with the PTUSE processor\cite{Bailes+20}, with the 10.5\,hour pilot survey conducted on 2019 October 12–13, and the orbital campaign on 2019 December 26-31 with nearly 12\,hours per day, at 4096 frequency channels and 8\,second sub-integrations. 

\subsection{Forming and processing the dynamic spectra}

The data were reduced using \textsc{psrchive}\cite{vanStraten+12} to 64 bins across pulse phase, and a standard template was formed by integrating all observations over time, frequency, and polarisation. Radio-frequency interference (RFI) was identified and removed using a version of the \textsc{CoastGuard} package\cite{Lazarus+16}, which was modified to use our standard template to determine an off-pulse region on which the outlier-detection statistics are performed. On average, 12.8\% of the band is affected by RFI\cite{Bailes+20}. The dynamic spectrum, $S(t, f)$ was produced using the \textsc{psrflux} function in \textsc{psrchive}, which fits the template to the observed pulse profiles at each sample in time, for each frequency channel, to recover the flux density.

The dynamic spectra were analysed primarily using the \textsc{scintools} (\url{https://github.com/danielreardon/scintools}) package\cite{Reardon+20, scintools}. Each dynamic spectrum was cropped at a maximum frequency of 1680 MHz, as the sensitivity of the MeerKAT band drops steeply at higher frequencies. The RFI-excised regions were first inpainted (refilled) using biharmonic functions\cite{Damelin+17} to reduce artefacts in the power spectrum. These inpainted dynamic spectra were used as an initial guess for generating a Wiener-filtered version of the spectrum\cite{Lin+21}. We then replaced the inpainted regions with the Wiener-filtered data to accurately repair the RFI-affected regions. 

Striations in the dynamic spectra, with both time and frequency (caused for example by gain variations, bandpass sensitivity changes, and intrinsic pulse jitter) were measured and removed by dividing the dynamic spectrum by the first mode of its singular value decomposition. A dynamic spectrum, processed in this way, is shown in Figure 1 of the main text, and shows the underlying fine-scale structure due to scintillation. The dynamic spectrum was then resampled with linear interpolation to have equal steps in wavelength, which has the effect of removing the frequency-dependence on arc curvatures\cite{Reardon+20}. This transformation would also make any inverted arclets more diffuse\cite{Sprenger+21}, but in the present data set there is no evidence for structure along the arcs. Finally, the mean flux density was subtracted and a Hanning window was applied to the outer 10 percent of the spectrum (in both time and frequency) to mitigate side-lobe response\cite{Reardon+20} in the power spectrum, yielding the processed dynamic spectrum, $\tilde{S}(t, \lambda)$.

\subsection{Forming and processing the secondary spectra}

The secondary spectrum for each observation was calculated as the squared magnitude of the zero-padded (to the nearest power of 2) two dimensional Fourier transform of the processed dynamic spectrum, $P(f_t, f_\lambda) = 10\log_{10}(|\mathcal{F}(\Tilde{S}(t, \lambda))|^2)$, where $\mathcal{F}$ indicates the Fourier transform. The axes of the spectrum are a differential Doppler shift $f_t$ (Fourier conjugate of time; x-axis) and a wavenumber $f_\lambda$ (Fourier conjugate of wavelength; y-axis) that corresponds to the more common differential time delay, $f_\nu \equiv \tau$, in a transformed frequency-time dynamic spectrum. The Nyquist limits for $f_\lambda$ and $f_t$ are, respectively, $11478$\,m$^{-1}$ and $62.5$\,mHz. The dynamic spectrum transformation to equal steps in wavelength induces a response in the secondary spectrum, which we correct by dividing the spectrum by the average power distribution as a function of $f_\lambda$ in the outer 10\% of the spectrum in $f_t$, which is far from the power of the arcs. In practice, since the secondary spectrum is in units of dB, we subtract the average dB power, with the result being a flat noise baseline with zero mean (e.g., white noise level, $w_{\rm noise}=0$\,dB) as a function of $f_\lambda$. For the secondary spectrum in Figure 2, the standard deviation of the off-arc white noise is $\sigma_{\rm noise}=5.6$\,dB. The colour scale for this figure ranges from $9$\,dB below the median power, to saturation at $15$\,dB below the maximum (for a total range of $-7.5$\,dB to $54$\,dB). We defined the colour scale relative to the median and maximum power to achieve a uniform visual contrast for all secondary spectra, which is used for an animation of the scintillation arcs available in the supplementary material.

The secondary spectrum shows parabolic scintillation arcs with curvatures $\eta$ defined with $f_\lambda = \eta f_t^2$ that reveal the geometry of thin scattering screens. The curvature depends on the fractional distance to the scattering region from the pulsar, $s$ ($s=0$ at the pulsar and $s=1$ at the observer) and the inverse square of the effective velocity $\mathbf{V}_{\rm eff}$, with
\begin{equation}
    \eta = \frac{D_{\rm psr}s(1-s)}{2\mathbf{V}_{\rm eff}^2\cos^2\psi},
\end{equation}
where $D_{\rm psr}$ is the pulsar distance and $\psi$ is the angle between the major axis of anisotropy of the scattered image and the direction of $\mathbf{V}_{\rm eff}$ (for an isotropic scattered image, $\cos^2\psi = 1$). The effective velocity through the scattering screen is,
\begin{equation}
\mathbf{V}_{\rm eff} = s\mathbf{V}_{\rm E} + (1-s)\mathbf{V}_{\rm p} - \mathbf{V}_{\rm IISM},
\end{equation}
where the velocities of the observer, the IISM, and the pulsar are $\mathbf{V}_{\rm E}$, $\mathbf{V}_{\rm IISM}$, and $\mathbf{V}_{\rm p}$, respectively. In this work, we analyse the power in the secondary spectrum as a function of $W = (2 \eta)^{-1/2}$, which is a convenient variable because it is proportional to $\mathbf{V}_{\rm eff}$.

Since PSR~J0437$-$4715 is observed near the regime of weak scintillation (with flux density modulation of typically 20\%), the scintillation arcs are observed to be much sharper than those observed for many other pulsars, where one sees not just interference between scattered rays and the direct line of sight but also between scattered rays themselves. This causes the arcs to broaden and inverted arclets to appear\cite{Cordes+06,Stinebring+22, Main+23}. However, the change in the direction and magnitude of the pulsar orbital velocity during $\sim 12$\,hour observations is significant enough to smear the arcs in the spectrum. To minimise the effect of this smearing on our curvature measurements, we split each observation into three dynamic spectra with $\sim 4\,$hour lengths. The secondary spectrum was computed for each of these segments and the arc curvatures were measured in each of them. Reducing the time span of the dynamic spectrum in this way enabled accurate curvature measurements, but reduced our resolution in $f_t$ and the signal-to-noise ratio of the arcs.

The normalised secondary spectrum\cite{Reardon+20} is a transformation in which parabolic arcs in $P(f_t, f_\lambda)$ are remapped to straight lines (by linear interpolation) with respect to some reference arc curvature, $\eta_{\rm ref}$. The x-axis of the normalised spectrum is then $f_{t,n} = f_t / f_{\rm arc}$, where $f_{\rm arc}^2 = f_\lambda / \eta_{\rm ref}$. However, for this work, the choice of $\eta_{\rm ref}$ in the transformation is arbitrary because the normalised x-axis is converted back into physical units using $W_{\rm ref} = \pm(2 \eta_{\rm ref})^{-1/2}$, yielding $P(W, f_\lambda)$. The spectrum is averaged along the $f_\lambda$ axis, to some maximum value $f_{\lambda, \rm max}$, to obtain the mean power as a function of $W$, $P(W | f_{\lambda, \rm max})$, where $|$ indicates that the average is taken for a given value of $f_{\lambda, \rm max}$. Distinct scintillation arcs in the secondary spectrum, $P(f_t, f_\lambda)$, correspond to peaks in $P(W | f_{\lambda, \rm max})$.

\subsection{Identifying scintillation arcs}
Each observation clearly shows myriad scintillation arcs in the secondary spectrum. In order to estimate the properties of the scattering screens from which these arcs originate, we measure and model the arc curvature variations over the pulsar orbit and between the long observations. To do this, distinct arcs were identified as peaks in $P(W | f_{\lambda, \rm max})$ that vary with time, approximately sinusoidally, since $W(t) \propto \mathbf{V}_{\rm eff}(t)$. We therefore represent the reduced data from the six day observing campaign as the mean power as a function of $W$ and time $t$, $P(W, t | f_{\lambda, \rm max} )$. The value of $f_{\lambda, \rm max}$ was chosen, for each arc, to be approximately the maximum extent of the arc in the secondary spectrum.

The peaks in $P(W | f_{\lambda, \rm max})$ were enhanced (i.e., increased contrast) by subtracting an estimate of the background power $B(W)$. This background can originate, for example, from scattered images that are not completely anisotropic, nearby bright arcs that contribute power across a range of $W$ because the pulsar is not observed in infinitely-weak scintillation\cite{Cordes+06, Reardon+20}, or from a background of scattering from unresolved screens or a uniform medium. We estimate $B(W)$ using a minimum filter, which is subsequently smoothed using a first-order Savitzky–Golay filter\cite{Savitzky+64} of the same size. The minimum filter, applied to the array $P(W | f_{\lambda, \rm max})$, returns the minimum value within a window of size $\mathcal{W}$ around each element, i.e., $B(W) = \min\limits_{w \in \mathcal{W}} P(W + w | f_{\lambda, \rm max})$. An appropriate window size for such a filter depends on the width of the arc (in $W$, with broad arcs requiring a larger windows), but should always be larger than this width. 

We also correct for small differential shifts in the location of peaks between the two halves of the spectrum (positive and negative $W$). Such shifts, $\Delta W$, arise because the main pulsar image has a small angular displacement away from the direct line-of-sight, caused by refraction. This displaces the apexes of the scintillation arcs from the origin\cite{Cordes+06}, which causes a small shift in apparent curvature (or $W$) in the normalised secondary spectrum (we derive and further discuss this shift in a subsequent section). The magnitude of $\Delta W$ varies stochastically on a short ($\sim$hour) timescale and is therefore different for each observation. The magnitude is also different for each arc because it depends on the location and velocity of the scattering screen. We correct for the shift, for each arc, by fitting for the value of $\Delta W$ that maximises the Pearson correlation coefficient for the two shifted halves, $P(W + \Delta W/2 | f_{\lambda, \rm max})$ for $W>0$, and $P(-(W - \Delta W/2) | f_{\lambda, \rm max})$ for $W<0$, using values of $W$ local to the arc (e.g., $\pm 10$\% of the arc $W$). The two shift-corrected halves of the background-subtracted $P(W, t | f_{\lambda, \rm max})$ are then averaged to increase the arc signal-to-noise, yielding a final reduced representation of the data, $P_{\rm arc}(W, t) = P({\rm abs}(W), t | f_{\lambda, \rm max}, \Delta W)$, that is optimised for each arc to show the curvature variations over the orbital campaign.

An example of $P_{\rm arc}(W, t)$ is shown in Supplementary Figure 1, with time replaced by binary orbital phase. Individual power enhancements corresponding to scintillation arcs are traced approximately by sinusoidal dashed white lines. The figure is divided into two halves, separated by a red line, where the data were reduced independently with different values of $f_{\lambda, \rm max}$, $\mathcal{W}$, and $\Delta W$. It is important to note that the values of these data reduction parameters were chosen to simultaneously show 12 arcs changing in curvature with time, but they do not optimise the power enhancements for any individual arc. For example, relative to the four bright arcs in the left half of the figure, there is a distorted faint arc (second from the left), which (in the secondary spectrum) extends to a lower $f_{\lambda, \rm max}$, has a broader distribution of power in $W$ (requiring larger $\mathcal{W}$), and is nearer to the pulsar (which affects $\Delta W$).

We identify distinct scintillation arcs as local maxima in $P_{\rm arc}(W, t)$ that are at least $2\sigma$ greater than neighbouring values, where $\sigma$ is the rms noise, which is a recent standard practice\cite{Reardon+20, Mckee+22, Ocker+24}. We additionally require that these maxima persist for at least $t=1$\,day, during which time the peak in $W$ evolves proportional to $\mathbf{V}_{\rm eff}\cos\psi$, which allows us to estimate the screen parameters with better accuracy than an arc detection at a single epoch. 

\subsection{Measuring the arc curvatures}
A heuristic sinusoid model of the form $H(t) = A \sin(\omega t + \phi) + C$ was fit to each arc in this data representation by tuning the free parameters $A$, $\phi$, and $C$ (dashed lines in Supplementary Figure 1). This model is an appropriate approximation for our data because the variations in $\mathbf{V}_{\rm eff}$ over the six day observing campaign are dominated by the pulsar motion in the near-circular orbit. Variations in $\psi$ are observed to be small because they are dominated by the pulsar proper motion and $\mathbf{V}_{\rm IISM}$. The differences between these heuristic models and our final physical models are $\mathcal{O}(1)$\%.

For each observation of each arc, this heuristic model provided a range in $W$ ($H(t) \pm 10$\%) in which to identify the peak in the background-subtracted $P(W | f_{\lambda, \rm max})$ distribution that is associated with the arc. We fit an inverse parabola\cite{Reardon+20} to the peak to obtain a measurement of $W$ and an estimate of the uncertainty, $\sigma_{\rm fit}$, to which we add in quadrature a constant that accounts for systematic errors, $\sigma_0=2$\,km\,s$^{-1}$\,$\sqrt{\rm kpc}$. This $\sigma_0$ value was the minimum observed rms residual for any arc following an initial curvature model fit with $\sigma_0=0$. In this way, we measured the local maxima of power in each observation, corresponding to each arc. These $W$ measurements and the uncertainties were the basis of our models. Overall, this process to identify and measure the arcs was interactive and was necessarily tuned for each screen. We have made the interactive tool and the chosen parameters for each screen available online, to facilitate reproducibility (see data availability statement). The measured $W$ values for each of the 25 arcs are shown in Figure 3, along with the models that are described in the following section.

\subsection{Modelling curvature variations}
We used the observations from both the orbital campaign and (when possible) the pilot survey, to model the arc curvature variations. The pilot survey was taken two months prior to the campaign, and therefore had a significantly different $\mathbf{V}_{\rm E}$ imprinted in the effective velocity. For the most precisely measured arcs that are near to the middle of the line-of-sight, we also detect significant variations in $\mathbf{V}_{\rm E}$ from a single orbit of the pulsar. We use the tools in \textsc{Astropy}\cite{astropy} and \textsc{scintools} to account for this component, while the pulsar proper motion and orbital velocity are computed from the pulsar timing ephemeris\cite{Reardon+16} and \textsc{scintools}. For each arc, we use an initial model obtained from the orbital campaign alone to predict the value of $W$ for the arc in the earlier pilot survey, after accounting for the change in $\mathbf{V}_{\rm E}$. If the arc was distinct in the earlier epoch, $W$ was measured and included in the model. We found that 11 arcs have distinct peaks in the earlier pilot survey consistent with their model and we used them to estimate the pulsar distance, as described in the following section.

The three unknown parameters in $\mathbf{V}_{\rm eff}$ for each screen are $s$ and two that describe the velocity of the plasma in the screen. If the scattered image is isotropic, we measure the plasma velocity in right ascension $V_{{\rm IISM},\alpha}$ and in declination $V_{{\rm IISM},\delta}$. If the scattered image is anisotropic, then the arc curvature only depends on the components along the major axis of anisotropy. The anisotropy need not be extreme (e.g. one dimensional or filamentary) as the model could be appropriate for elongated images with axial ratios $A_r \gtrsim 2$\cite{Reardon+20}. In this case, we measure the plasma velocity along the axis of anisotropy $V_{{\rm IISM},\zeta}$, and the angle of the scattered image on the sky $\zeta$, measured East of North. To model each measurement, we compute $\psi$ as the angle between $\zeta$ and the velocity vector $\mathbf{V}_{\rm eff}$ at the measurement epoch.

Bayesian inference is used to infer posterior probability distributions for the model parameters, given the measured data (see\cite{Thrane+19} for an overview). We use \textsc{Bilby}\cite{Ashton+19} and \textsc{dynesty}\cite{Speagle20} to perform parameter inference and model selection through the Bayes factors, $\mathcal{B}$. We assume a Gaussian likelihood function for our measurements of $W$ and include two white noise parameters, for each screen, to modify the formal measurement uncertainties\cite{Walker+22, Askew+23}, which ensures residuals with reduced chi-squared values near unity and more reliable parameter uncertainties. The modified uncertainties are $\sigma_{\rm new} = \sqrt{(F\sigma_{\rm fit})^2 + Q^2}$, where $F$ and $Q$ are free parameters and inferred jointly with the parameters of each screen. These white noise parameters are required because $\sigma_{\rm fit}$ underestimates the true uncertainty owing to several effects: the shape of the peak power is not a parabola\cite{Reardon+20}, the peak can be biased by our background subtraction or the presence of other arcs nearby in curvature, and small phase gradients and random fluctuations of the ISM velocity or anisotropy angle jitter the curvatures\cite{Askew+23}.

The largest source of uncertainty among the known timing model parameters of PSR~J0437$-$4715 is the longitude of ascending node $\Omega$, which is measured from small distortions of the projected orbit due to annual-orbital parallax\cite{Kopeikin95}. To include the uncertainty of this parameter in our model, we use a conservative Gaussian prior probability distribution with mean and standard deviation $\bar{\Omega} = 207^\circ$ and $\sigma_{\Omega} = 2.4^\circ$ respectively, corresponding to the 95\% confidence interval from a pulsar timing model\cite{Reardon+16}. The remaining timing model parameters are fixed at their known values and we adopt broad uniform priors for the scattering screen parameters. For pulsars where $\Omega$ is unknown, it can be inferred from the scintillation arcs. However, longer time baselines that capture the changing $\mathbf{V}_{\rm E}$\cite{Reardon+20} will be required for precise measurements.

For each screen we fit models for isotropic and anisotropic scattered images (Table 1). If a model is favoured with $|\ln{\mathcal{B}}| \geq 8$, this can be interpreted as strong supporting evidence for one model over the other.

\subsection{Screen and pulsar distances}
The measured locations of the ISM screens, relative to the Earth, pulsar, and the Local Bubble, are shown in Extended Data Figure 3. As discussed in the main text, the location of these screens is impacted by selection effects; notably a bias for screens with higher $\mathbf{V}_{\rm eff}$. As a result, the observed screens will appear preferentially near to the pulsar and with a velocity in the direction opposite to the pulsar proper motion. Only three arcs are observed at higher curvature than a hypothetical screen placed at $s=0.5$ with stationary velocity with respect to the Solar system barycentre. However, the apparent higher density of screens towards the edge of the Bubble may also have a physical component, as the plasma density of the Bubble could increase towards the edge in the direction of the pulsar. The apparent alignment of the inferred anisotropy orientations may also have a physical component, as the screen plasma structures may be preferentially aligned with the larger-scale magnetic
field of the Galaxy.

While the pulsar distance was fixed for our arc curvature models, we also demonstrate that it can be inferred using our measurements. For 11 of the screens, the arcs were pronounced enough to be unambiguously identified in the early pilot survey observation as well as the orbital campaign that followed two months later. These screens can each constrain the pulsar distance and a combined posterior distance constraint using all screens is taken as the product of the likelihood distributions from individual screens, since the data are independent and we assume a uniform distance prior. The constraints from individual screens, and their product is shown in Supplementary Figure 2.

One scintillation arc (ID 2 in Table 1), dominates the constraint on the pulsar distance (blue line Supplementary Figure 2). This arc is one of the brightest in our sample and it is also isolated from the other arcs in curvature space, making the measurements of $W$ both precise and accurate (e.g. free of systematics caused by nearby bright arcs). This arc is also sensitive to the longitude of ascending node of the pulsar orbit, $\Omega$, which is a quantity that is usually not known \textit{a priori} for most pulsar systems. Inferring $\Omega$ with a uniform prior over 360 degrees yields a unique solution, $\Omega=204^{+10}_{-13}$ (degrees, East of North), which is consistent with the more precise value from pulsar timing. The posterior probability distribution for $\Omega$ and $D_{\rm psr}$ from this scintillation arc is shown in Extended Data Figure 1. This pulsar distance constraint, $D_{\rm psr}=158^{+3}_{-6}$\,pc, is more representative of what might be possible for more pulsar systems with precision scintillation arc measurements.

\subsection{Interpreting the bow shock}

The bow shock of PSR~J0437$-$4715 was first observed in H$\alpha$ in 1995\cite{Fruchter95}. Here we have used three H$\alpha$ images from SOAR of 600\,s integration time each, observed in 2012\cite{Brownsberger+14}. These images were processed by taking the median intensity across the three days, removing artefacts in the image due to cosmic rays (following \citet{Brownsberger et al. (2014)}{Brownsberger+14}), and Gaussian smoothing by two pixels ($0.308\,$arcseconds). The pulsar position was taken to be the flux centroid of the point source associated with the white-dwarf companion. Using the precise distance to the pulsar, the coordinates were then converted to distance in au, centred on the pulsar.

\citet{Wilkin (1996)}{Wilkin96} gives an analytical expression for the shape of the contact discontinuity as a function of position angle $\theta$ (from the pulsar at the origin), which is behind the H$\alpha$ emitting forward shock
\begin{equation}
    r_{cd} = r_0 (3(1 - \theta\cot\theta))^{1/2} / \sin{\theta},
\end{equation}
where $r_{cd}$ is the distance to the contact discontinuity and $r_0$ is the standoff radius. This equation assumes that the pulsar wind is close to isotropic and that the ambient medium has uniform density, and ignores post-shock pressure. The model agrees well with the shock head for PSR~J0437$-$4715. However, post-shock pressure causes the observations to depart from the model in the tail. Since our measured radial distance to the shock is behind the shock head, we instead adopt a more general model of the shock shape, which traces the H$\alpha$ emission well for $\sim 2000$\,au (beyond the pulsar) before departing in the tail of the shock. We fit this model simultaneously to the H$\alpha$ emission and our measured radial distance. 

We begin by describing the radial distance to the H$\alpha$-emitting shock, $r_s$, with a conic section curve
\begin{equation}
    r_s = \frac{p} { 1 + e \cos{\theta} }
\end{equation}
where $e$ is the curve eccentricity and $p = D_x e$, where $D_x$ is the directrix distance (related to the shock vertex distance). This curve is rotated from $\phi=0$ to $\phi=\pi$ radians about an axis pointing from the pulsar towards the observer, to form a three dimensional cone, which we describe in Cartesian coordinates
\begin{align}
    x_0 = &  \,r_s \sin{\theta} \cos{\phi} \notag \\ 
    y_0 = &  \,r_s \sin{\theta} \sin{\phi} \\
    z_0 = &  \,r_s \cos{\theta} \notag
\end{align}
The cone model is then rotated by inclination angle $i$,
\begin{align}
    x_i = &  \, x_0 \cos{i} + z_0 \sin{\phi} \notag\\
    y_i = &  \, y_0 \\
    z_i = &  \, -x_0 \sin{i} + z_0 \cos{i} \notag
\end{align}
and finally by a position angle $\beta$, which is expected to be close to that of the proper motion vector
\begin{align}
    x = &  \, - x_i \cos{\beta} + y_i \sin{\beta} \notag\\
    y = &  \, x_i \sin{\beta} + y_i \cos{\beta} \\
    z = &  \, z_i \notag
\end{align}
where we have defined $y$ to be in the declination direction and $x$ to be right ascension. The scattering screen is then located in the $z$ direction. The free parameters in the model are $e$, $\beta$, $i$, $p$, and a white noise parameter that accounts for errors in our measurements of the H$\alpha$ peak intensity (described below). The geometric model appropriate for the H$\alpha$ emission and the scattering screen is assumed to be the projected outer edge of this shape, which is computed numerically as the convex hull\cite{convex_hull} of a set of points determined by a grid in $\phi$ and $\theta$.

For a range of angles from the pulsar proper motion direction (e.g., $\theta_{\rm i} = \pm 60^\circ$), we measured the intensity as a function of distance $r$ from the pulsar, $I(\theta_i, r)$.  This H$\alpha$ intensity was fit with a Gaussian function and the location of the peak was taken to be the centre of the $H\alpha$ emission region, measured at a projected distance to the vertex $1445 \pm 6\,$au from the pulsar. Posterior probability distributions for the model parameters were inferred using Bilby, assuming a Gaussian likelihood and with the standard error in the H$\alpha$ peak measurements as a free white noise parameter. We find that the maximum likelihood model describes a hyperboloid with $e = 1.06 \pm 0.02$, $p = 2770\pm60 \,$au, tilted by $i = 113 \pm 3^\circ$, and offset from the proper motion by $4.3^\circ$, at $\beta = 124.8 \pm 0.6^\circ$ (East of North).

\subsection{Measuring arc power distributions}
While the arc curvatures reveal the distance and velocity of each scattering screen, the power along and interior to the arcs contains information about the (transverse) spatial distribution of the scattering plasma. For each arc, we determine the spectral index of the power decay with larger $f_\lambda$, the maximum detectable scattering delay (with corresponding maximum scattering angle), and the magnitude of a phase gradient caused by refraction.

Following \citet{Reardon et al. (2020)}{Reardon+20}, the arc delay profile $\mathcal{D}(f_{\lambda})$ in the secondary spectrum was scaled such that spectral index $\gamma$ would correspond to the index of the phase spectrum (where $\gamma = 11/3$ for Kolmogorov turbulence). We fit for the spectral exponent $\gamma$ and amplitude $K$ in the relation

\begin{equation}
  f_{\lambda}^{1/2}\mathcal{D}(f_\lambda) = K f_{\lambda}^{-\gamma/2}.
\end{equation}
We define the maximum extent of each arc as the scattering time delay $\tau_{\rm max}$ beyond which  $K f_{\lambda}^{-\gamma/2} < w_{\rm noise}$. The measured $\gamma$ and $\tau_{\rm max}$ for each arc is listed with the arc curvature parameters in Table 1. 

The maximum detectable scattering angle (angle of deflection) from each screen, corresponding to each $\tau_{\rm max}$, is\cite{Lee+75}
\begin{equation}
\label{eqn:alpha}
    \alpha_{\rm max} = \sqrt{\frac{2c\tau_{\rm max}}{D_{\rm psr}s(1-s)}}.
\end{equation}
We have calculated this deflection angle for each arc and find that the interstellar plasma screens have similar magnitudes in scattering strength, though screens nearer to the pulsar show larger $\alpha_{\rm max}$. The mean and standard deviation from the ISM screens is $\alpha_{\rm max, ISM} = 4.8 \pm 0.8\,$mas. The implication is that the arcs that extend furthest in $f_\lambda$ (or equivalently, $\tau$) are not dominant in scattering strength but are enhanced because they are nearer to the centre of the line-of-sight ($s=0.5$). However, the bow shock arcs clearly originate from significantly stronger scattering, with $\alpha_{\rm max}$ up to an order of magnitude greater than that of the ISM screens, as shown in Extended Data Figure 2.

\subsection{The effect of refraction}
In the case of weak scintillation, the main pulsar image is much brighter than the surrounding diffuse scattered image. This results in a sharp parabolic arc, or multiple non-interfering parabolae for multiple screens, as is the case for our PSR~J0437$-$4715 observations. Phase gradients can lead to refractive angular shifts of the main pulsar image as well as the scattered images. This is observed through tilts of auto-covariance functions\cite{Reardon+23}, or equivalently a shift in the apex of the scintillation arc and an asymmetry in arc brightness about the $f_{t}=0$ axis\cite{Cordes+06}. The location of the apex shifts from the origin because the bright main image is offset from the direct line of sight. This allows part of the scattered image to take the direct path and arrive at the observatory before the main image, giving a negative differential delay for these scattered waves.

For a refractive angular shift $\theta_R$, the new apex will be centered on 
\begin{equation}
f_{t, R} = \frac{\theta_R \mathbf{V}_{\rm eff} \cos(\psi)}{s \lambda} \quad \textrm{and} \quad  f_{\lambda, R} = \frac{\theta_R^2 D_{\rm psr}(1-s) }{2 s \lambda^2}.
\end{equation}
Here we calculate the effect of such a refractive shift on our inferred values of $W$.  The normalised secondary spectra remap parabolic arcs to vertical lines of constant $W$ assuming that the apex of the parabolae lie at the origin. If the true value of $W$ is $W_0 \equiv f_{t,0}/\sqrt{2 f_{\lambda, 0}}$, the measured values of $W$ from the left ($f_{t}<0$) and right ($f_{t}>0$) halves of the secondary spectrum will be, respectively,
\begin{equation}
W_{l,\rm meas} = \frac{f_{t,0} - f_{t,R}}{\sqrt{2(f_{\lambda, 0} - f_{\lambda, R})}} \quad \textrm{and} \quad  
W_{r,\rm meas} = \frac{f_{t,0} + f_{t,R}}{\sqrt{2(f_{\lambda, 0} - f_{\lambda, R})}}.
\end{equation}
The refractive shifts observed in PSR~J0437$-$4715 are much less than the arc extents and $\theta_R \ll 1$. The shift in $f_t$ is linear in $\theta_R$, while $f_{\lambda, R} \propto \theta_R^2$, so a shift in $f_t$ is the dominant effect for our observations.

Our final measure of $W$ is the average of separate measurements from the left and right halves of the spectrum, $W_{l,\rm meas}$ and $W_{r,\rm meas}$, respectively,
\begin{equation}
\label{eqn:avg}
    W_{\rm meas} = \frac{f_{t,0}}{\sqrt{2(f_{\lambda, 0} - f_{\lambda, R})}} \approx  \frac{f_{t,0}}{\sqrt{2f_{\lambda, 0} }} = W_{0}.
\end{equation}
The difference between $W_{l,\rm meas}$ and $W_{r,\rm meas}$ is a measure of the refractive shift, through 
\begin{equation}
\label{eqn:phase_grad}
    \Delta W_{\rm meas} = \frac{2f_{t,R}}{\sqrt{2(f_{\lambda, 0} - f_{\lambda, R})}} \approx \frac{2f_{t,R}}{\sqrt{2f_{\lambda, 0}}} = \frac{2\theta_R \mathbf{V}_{\rm eff} \cos{\psi}}{\bar{\theta} \sqrt{D_{\rm psr}s(1-s)}} ,
\end{equation}
where $\bar{\theta}$ is a weighted mean scattering angle for each screen that arises by averaging the normalised secondary spectrum over $f_\lambda$ with weights given by the Kolmogorov power spectrum\cite{Reardon+20}. We estimate $\bar{\theta}$ for each arc using our $\tau_{\rm max}$ and solve for $\theta_R$, since our curvature models give $\mathbf{V}_{\rm eff}$ and $s$ for each screen. The observed $\Delta W_{\rm meas}$ are larger than the errors in $W$, and only through averaging both sides of the arc (Equation \ref{eqn:avg}) do we achieve precise results.

We find that our measured offsets $\Delta W_{\rm meas}$ are consistent with small refractive shifts of order $\theta_R \sim \mathcal{O}(0.1)\,$mas occurring somewhere on the line of sight. The magnitude of $\Delta W_{\rm meas}$ in each arc, resulting from this refractive shift, depends on the screen distance and the extent of the arc in $f_\lambda$, following Equation \ref{eqn:phase_grad}. We observe that these small angular displacements have a consistent average direction but vary in magnitude on short ($\lesssim 1\,$hour) timescales, which makes them challenging to measure and interpret. However, the small timescale implies an origin near the observer or pulsar, though there is no clear variability with pulsar orbital phase. We attribute these small refractive angular offsets to the ionosphere.

\end{methods}

\begin{addendum}
 \item[Correspondence] Correspondence and requests for materials
should be addressed to D.J.R.~(email: dreardon@swin.edu.au).
 \item[Data availability] The dynamic spectra from our observations will be available, upon publication, from figshare at \url{doi.org/10.6084/m9.figshare.27311715}. Any further data may be provided upon reasonable request to the corresponding author. 
 \item[Code availability] Code for processing the dynamic spectra and the interactive tools for arc identification and curvature fitting are available at \url{github.com/danielreardon/J0437-Scintillation-arcs}, including the parameters used to model screens for this work.
 \item  The MeerKAT telescope is operated by the South African Radio Astronomy Observatory, which is a facility of the National Research Foundation, an agency of the Department of Science and Innovation. Part of this work was undertaken as part of the Australian Research Council Centre of Excellence for Gravitational Wave Discovery (project numbers CE170100004 and CE230100016). RMS acknowledges support through Australian Research Council Future Fellowship FT190100155. AP acknowledges financial support from the European Research Council (ERC) starting grant 'GIGA' (grant agreement number: 101116134) and through the NWO-I Veni fellowship.
\item[Author Contributions] DJR devised the project, led the data analysis, and drafted the manuscript. RM contributed significantly to the analysis of the scintillation arcs and bow shock and assisted with manuscript preparation. SKO contributed significantly to the interpretation of the scintillation arcs and assisted with manuscript preparation. RMS and MB were involved in observation planning and provided input on the manuscript. DJR, RMS, MB, FC, MG, AJ, MK, AP, RS, WvS, and VVK contributed to the foundational work of the MeerTime project, which facilitated the collection of this dataset under the MeerKAT Pulsar Timing Array (MPTA) program. All authors reviewed and provided feedback on the manuscript.
\item[Competing Interests] The authors declare that they have no competing financial interests.
\end{addendum}


\renewcommand{\figurename}{Supplementary Figure} 
\setcounter{figure}{0}

\begin{figure}
\centering
  \includegraphics[width=1\textwidth]{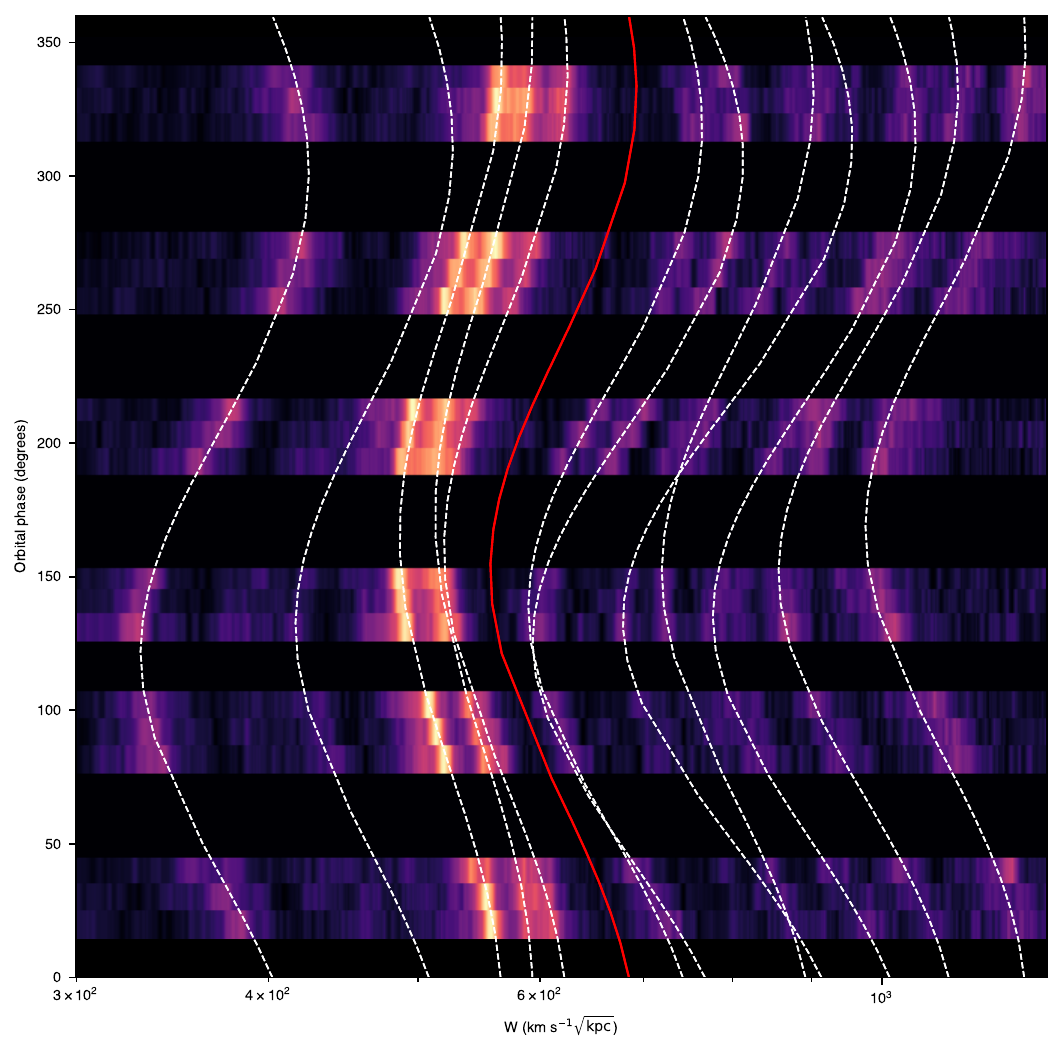}
  \caption{Orbital variations of the scintillation arcs. This summed, background subtracted, normalised secondary spectrum, $P_{\rm arc}(W, t)$ (see methdos), shows the arc variations in $W$ caused by the orbital velocity of the pulsar. Heuristic models for twelve of the observed screens are shown with dashed white lines. Only a segment in $W$ of the full secondary spectrum is shown, to focus on the clearest arcs. The red line separates two regions where the data reduction was different, to improve clarity of the arcs. In the left (right) region, the spectra are summed to a maximum value of $f_{\lambda, \rm max} = 11696$\,m$^{-1}$ ($2924$\,m$^{-1}$) and the background power filter uses a window of size $\mathcal{W}=800\,$pixels ($400$\,pixels). The phase shift due to refraction, $\Delta W$, was fitted independently in each region and for each day (consisting of three observations).}
\label{fig:norm_sspec}
\end{figure}

\begin{figure}
\centering
  \includegraphics[width=.8\textwidth]{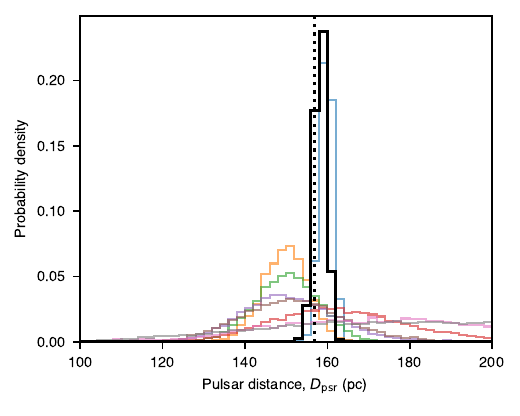}
  \caption{Probability density distribution for the pulsar distance, $D_{\rm psr}$, from scintillation arcs. Posterior distributions for $D_{\rm psr}$ from 11 individual screens are shown in colours, with the combined constraint shown in black. The vertical dotted line marks the precise distance from pulsar timing\cite{Reardon+24}. The most sensitive screen with the tightest constraint (blue) is labelled as ``ID 2" in Table 1.}
\label{fig:distance}
\end{figure}

\clearpage


\end{document}